\documentclass[fleqn,usenatbib]{mnras}
\usepackage{multirow}
\usepackage{rotating}
\usepackage{colortbl}
\usepackage{color}
\usepackage{xcolor}
\usepackage[fleqn]{amsmath}
\usepackage{amssymb}
\usepackage{amsfonts}
\usepackage{verbatim}
\usepackage{scalefnt}
\usepackage{lscape}
\usepackage{booktabs}
\usepackage{multicol}
\usepackage{newtxtext,newtxmath}
\usepackage{graphicx}	
\usepackage{amsmath}	
\usepackage[T1]{fontenc}

\DeclareRobustCommand{\VAN}[3]{#2}
\let\VANthebibliography\thebibliography
\def\thebibliography{\DeclareRobustCommand{\VAN}[3]{##3}\VANthebibliography}

\newcommand{\beq}{\begin{equation}}
\newcommand{\eeq}{\end{equation}}

\title[Recent JWST detections in light of cosmological simulations]{Is the James Webb Space Telescope detecting too many AGN candidates?}

\author[Melanie Habouzit]{Melanie Habouzit$^{1,2}$\thanks{E-mail: melanie.habouzit@unige.ch}\\
$^1$ Department of Astronomy, University of Geneva, Chemin d'Ecogia, CH-1290 Versoix, Switzerland\\
$^2$ Max-Planck-Institut f\"ur Astronomie, K\"onigstuhl 17, D-69117 Heidelberg, Germany\\
}

\date{2024}
\pubyear{2024}

\begin{document}
\label{firstpage}
\pagerange{\pageref{firstpage}--\pageref{lastpage}}
\maketitle

\begin{abstract}
In less than two years of operation, the James Webb Space Telescope (JWST) has already accelerated significantly our quest to identify active massive black holes (BHs) in the first billion years of the Universe's history. At the time of writing, about 50 AGN detections and candidates have been identified through spectroscopy, photometry, and/or morphology. Broad-line AGN are about a hundred times more numerous than the faint end of the UV-bright quasar population at $z\sim 5-6$. In this paper, we compare the observational constraints on the abundance of these AGN at $z\sim 5$ to the populations of AGN produced in large-scale cosmological simulations. 
Assuming a null fraction of obscured simulated AGN, we find that while some simulations produce more AGN than discovered so far, some others produce a similar abundance or even fewer AGN in the bolometric luminosity range probed by JWST. 
Keeping in mind the large uncertainty on the constraints, we discuss the implications for the theoretical modeling of BH formation and evolution in case similar constraints continue to accumulate.
At the redshift of interest, the simulated AGN populations diverge the most at $L_{\rm bol}\sim 10^{44}\, \rm erg/s$ (by more than a dex in the bolometric luminosity function). This regime is most affected by incompleteness in JWST surveys. However, it holds significant potential for constraining the physical processes determining the assembly of BHs (e.g., seeding, feedback from supernova and AGN) and the current abundance of broad-line AGN with $\geqslant 10^{44.5}\, \rm erg/s$.

\end{abstract}


\section{Introduction}

In the past decade, the community successfully studied galaxy formation in the cosmological context with large-scale cosmological hydrodynamical simulations of the Universe. Capturing highly non-linear physical processes, these simulations of $\sim 50-400\, \rm cMpc$ box side length have effectively shown that it is feasible to attain satisfactory agreements with observational constraints on galaxy properties over several Gyr of our Universe's history, including galaxy clustering, galaxy and halo stellar mass functions, galaxy star formation rates, morphologies, sizes, and color bimodality. 

Massive black holes (BHs) are a fundamental component of these simulations, as the main source of energy able to regulate star formation in massive galaxies and responsible for giving rise to the population of quiescent galaxies observed in the Universe possibly as early as $z\sim 5$ \citep[see][for a recent detection using JWST]{2024arXiv240405683D}. 

At the time the simulations were conceived, constraints on the BH population were limited and primarily focused on the local Universe, for example, on inactive BHs in the bulge of massive galaxies at $z=0$. The BH models (seeding, accretion, feedback, coalescence) used in the simulations remained unconstrained at high redshift. 
This lack of constraints resulted in significant variations in their populations, as summarized in \citet{2020arXiv200610094H,2022MNRAS.509.3015H}. 
However, collectively, these simulations enabled us to evaluate the current uncertainties on, for example, the theoretical BH mass function and AGN luminosity function, correlations between BH properties and those of the host galaxies, and more broadly, uncertainties in our subgrid modeling (e.g., modeling and efficiency of feedback processes, seeding, accretion).

In parallel, large-scale surveys conducted over the past decade have unveiled hundreds of UV-bright active BHs referred to as high-redshift quasars ($z\geqslant 4$) with bolometric luminosties of $L_{\rm bol}\geqslant 10^{46}\,\rm erg/s$. More recently, fainter quasars with $L_{\rm bol}= 10^{45}-10^{46}\,\rm erg/s$ (corresponding to $M_{\rm 1450}\sim -24$) have been successfully identified in moderately deep surveys, such as the Canada-France High-redshift Quasar Survey \citep[CFHQS, e.g.,][]{2007AJ....134.2435W,2009AJ....137.3541W,2010AJ....139..906W,2015ApJ...813L..35K} and the Subaru High-redshift Exploration of Low-Luminosity Quasars \citep[SHELLQs, e.g.,][]{2016ApJ...828...26M,2018ApJS..237....5M,2019ApJ...883..183M,2019ApJ...872L...2M}. These new systems have enabled the characterization of the faint end of the quasar luminosity function at $z=6$ \citep{2018ApJ...869..150M} and $z=5$ \citep[e.g.,][]{2013ApJ...768..105M,2013A&A...557A..78M,2018AJ....155..131M,2020ApJ...904...89N}, down to $M_{1450}<-22.32$. 

With JWST, it is now possible to identify at the same redshifts UV-faint AGN down to the bolometric luminosity range $L_{\rm bol}=10^{44}-10^{46}\, \rm ergs/s$ \citep[e.g.,][]{2023ApJ...942L..17O,2023ApJ...946L..14K,2023arXiv230512492M,fujimoto2023uncover,2023ApJ...952..142F,2023ApJ...957L...7K}.
Candidates are numerous, and come primarily from the identification of a broad component in the H${\rm \alpha}$ or H${\rm \beta}$ emission line (from 1000 km/s up to more than $3000$ km/s). A non-negligible fraction of the broad-line systems are quite heavily dust obscured with attenuation of up to $A_{\rm v}\sim 4$. Some candidates have a V-shape SED \citep{2023ApJ...946L..14K,2023arXiv230607320L,2024arXiv240108782P}, i.e. a steep slope in the rest-frame optical which could come from a reddened AGN or emission from dusty star formation, and an additional rest-frame UV component coming potentially from low-level star formation or scattered unobscured light from the AGN \citep{2023ApJ...946L..14K,2023arXiv230905714G,2006ApJ...642..673P}. Some other candidates are identified photometrically from SED fitting, their singular colors, and/or their compactness \citep[e.g.,][]{2023MNRAS.525.1353J}.  
Although many of the observed systems have strong evidence for being AGN (and would hardly be explained by other types of sources), we use the term ``candidates'' in the paper to be general and reflect the uncertainties on the properties of JWST broad-line AGN and other systems (e.g., $L_{\rm bol},\, M_{\rm BH}$). Uncertainties on $M_{\rm BH}$ include at least 0.5 dex, arising from the use of virial mass estimators based on the width of broad emission lines and broad-line region size estimates, which are calibrated using reverberation mapping of local AGN.

With similar results in different surveys, 
AGN candidates yield large number densities of $\sim 10^{-5}\, \rm cMpc^{-3}\, mag^{-1}$ in the redshift range $z\sim 4-6$ \citep[e.g.,][]{2023arXiv230605448M,2023arXiv230311946H,2023arXiv230905714G}.

The abundances of UV-faint AGN (and undetectable inactive BHs) at high redshift is crucial for constraining the origin(s) of BH seeds \citep[e.g., assessing the efficiency of their formation in the Universe, determining their initial mass distribution, see the recent reviews of][]{2020ARA&A..58...27I,2021NatRP...3..732V}, understanding their early growth (which is expected to be predominantly obscured in rest-frame UV for a significant fraction of their early evolution, a trend supported by JWST detections of reddened AGN), and investigating how they settle into a potential co-evolution pattern with their host galaxies.

Additionally, although there is a consensus that reionization is primarily triggered by ionizing photons escaping from the numerous low-mass galaxies \citep{2022arXiv221206177T}, the growing sample of JWST AGN candidates and their BH to stellar mass ratios may require to revisit the contribution of AGN. \citet{2024arXiv240111242D} recently showed that the UV-faint AGN would provide about $\sim 20\%$ of the total reionization budget (mostly from AGN in galaxies with $M_{\star}\geqslant 10^{9}\, \rm M_{\odot}$), and confirmed pre-JWST results that they would only contribute as significantly as galaxies at the late stage of reionization ($z\sim 6.2$).

In this paper, we provide a theoretical perspective on the abundance of AGN candidates unveiled with JWST by comparing their number to predictions from large-scale cosmological simulations. We describe the simulations in Section~\ref{sec:method}. We compare simulated and observed BH populations in the $M_{\rm BH}-M_{\star}$ diagram in Section~\ref{sec:diagram}, and their BH mass functions and AGN luminosity functions as well as potential explanations for the discrepancies with observations in Section~\ref{sec:abundance}. Finally, Section~\ref{sec:summary} summarise our findings.

\section{Cosmological simulations}
\label{sec:method}
We use nine large-scale cosmological hydrodynamical simulations: Illustris, TNG50 (highest resolution and smallest volume of all the simulations studied in the paper), TNG100, TNG300 (the largest volume and lowest resolution of the TNG suite), Horizon-AGN, EAGLE, SIMBA, BlueTides, and Astrid. 
All these simulations have volumes ranging from $50^{3}$ to ${\sim}400^{3}\, \rm cMpc^{3}$, dark-matter mass resolutions of $\sim$$5\times 10^{5}-8\times 10^{7}\, \rm M_{\odot}$, and spatial resolutions of 1-2 ckpc. All simulations run down to at least $z=4$ (our range of interest), except BlueTides (snapshots are available for $z\geqslant 6.5$).
The simulations model the evolution of the dark and baryonic matter contents in an expanding space-time, with different cosmologies (all consistent with WMAP or Planck). 
Physical processes taking place at small scales below the galactic scale, such as  star formation, feedback from supernovae (SNe), BH formation, growth, and feedback, are modeled as ``subgrid'' physics. These models vary from simulation to simulation \citep[][for a summary]{2020arXiv200610094H}. More detailed descriptions of the simulations and their BH modelling can be found in \citet{2014MNRAS.445..175G,2014MNRAS.444.1518V} for Illustris, \citet{2017arXiv170302970P,2018MNRAS.479.4056W} for TNG, \citet{2016MNRAS.463.3948D,2016MNRAS.460.2979V} for Horizon-AGN, \citet{2015MNRAS.446..521S,2016MNRAS.462..190R,2018MNRAS.481.3118M,2017MNRAS.468.3395M} for EAGLE, \citet{2019MNRAS.486.2827D,2019MNRAS.487.5764T,2020arXiv201011225T} for SIMBA, \citet{2016MNRAS.455.2778F} for BlueTides, and \citet{2022MNRAS.513..670N,2022MNRAS.512.3703B} for Astrid. \\

Cosmological simulations form BHs in galaxies under specific seeding conditions.  Depending on the simulations, BHs form in halos when their mass exceeds a certain threshold (e.g., $M_{\rm h}\geqslant 1.5, \, 7.1, \,7.2$ or $7.4\times 10^{10}\, \rm M_{\odot}$ for EAGLE, Illustris, BlueTides and TNG, respectively), in galaxies with $M_{\star}\geqslant 10^{9.5}\, \rm M_{\odot}$ (SIMBA), or through a combination of both conditions, with halo mass exceeding $M_{\rm h}\geqslant 7.4\times 10^{9}\, \rm M_{\odot}$ and stellar mass $M_{\star}\geqslant 3\times 10^{6}\, \rm M_{\odot}$ in the case of Astrid. Horizon-AGN employs a more physically motivated model, based on the local properties of the gas.
The number of BHs produced in the simulations is the result of the seeding prescriptions and the assembly of dark matter halos and galaxies. At high-redshift, the number of galaxies with $M_{\star}\geqslant 10^{9}\, \rm M_{\odot}$ hosting a BH is close to unity \citep{2022MNRAS.514.4912H}. For example at $z=5$, this corresponds to a number density of galaxies with at least one BH varying in the range $\sim 1.9-6.5 \times 10^{-4}\, \rm cMpc^{-3}$ for $M_{\star}\sim 10^{9}\, \rm M_{\odot}$ and $1.9-4.7\times 10^{-5}\, \rm cMpc^{-3}$ for $10^{10}\rm \, M_{\odot}$ (bin of 0.2 in log space), across the simulations.

The initial mass of the BHs is fixed in most simulations and varies from $10^{4}$ to $10^{6}\, \rm M_{\odot}$ across simulations. Unlike the other simulations presented here, in Astrid the initial mass of BHs follows a power-law distribution ranging from $M_{\rm seed}=4.4\times 10^{4}$ to $4.4\times 10^{5} \, \rm M_{\odot}$.
For the purposes of this paper, we focus solely on the most massive BH within each simulated galaxy. 

BHs accrete gas through the Bondi-Hoyle-Lyttleton formalism in all simulations except SIMBA in which an additional gravitational torque limited model is added for the cold gas. 
We compute the luminosity of the simulated AGN with $L_{\rm bol}=\epsilon_{\rm r}\dot{M}_{\rm BH} c^{2}$ and use a radiative efficiency of $0.1$ for all simulations. We also tested a model in which the luminosity is computed differently for radiative-efficient and radiatively inefficient AGN, but it does not impact our result in the regime of interest (only faint AGN at or below the knee of the luminosity function are impacted).

Interestingly, most of the simulations are not calibrated based on
the properties of their BH population, but rather on galaxy properties (e.g.,
mass function, size) The only exception is a rough agreement with the $M_{\rm BH}-M_{\rm star}$ diagram in the local Universe. Historically, BHs have primarily been included in simulations to solve galaxy quenching via AGN feedback. The subgrid models were not tuned to reproduce the observed constraints on the AGN luminosity function, which explains why the AGN luminosity functions diverge by orders of magnitude across different simulations. Since the efficiency of AGN feedback is often tuned in large-scale simulations, different simulations producing the same galaxy properties can produce different AGN luminosity functions and $M_{\rm BH}-M_{\rm star}$ diagrams.

\section{BH and galaxy stellar mass diagram}
\label{sec:diagram}
\begin{figure*}
    \includegraphics[scale=0.52]{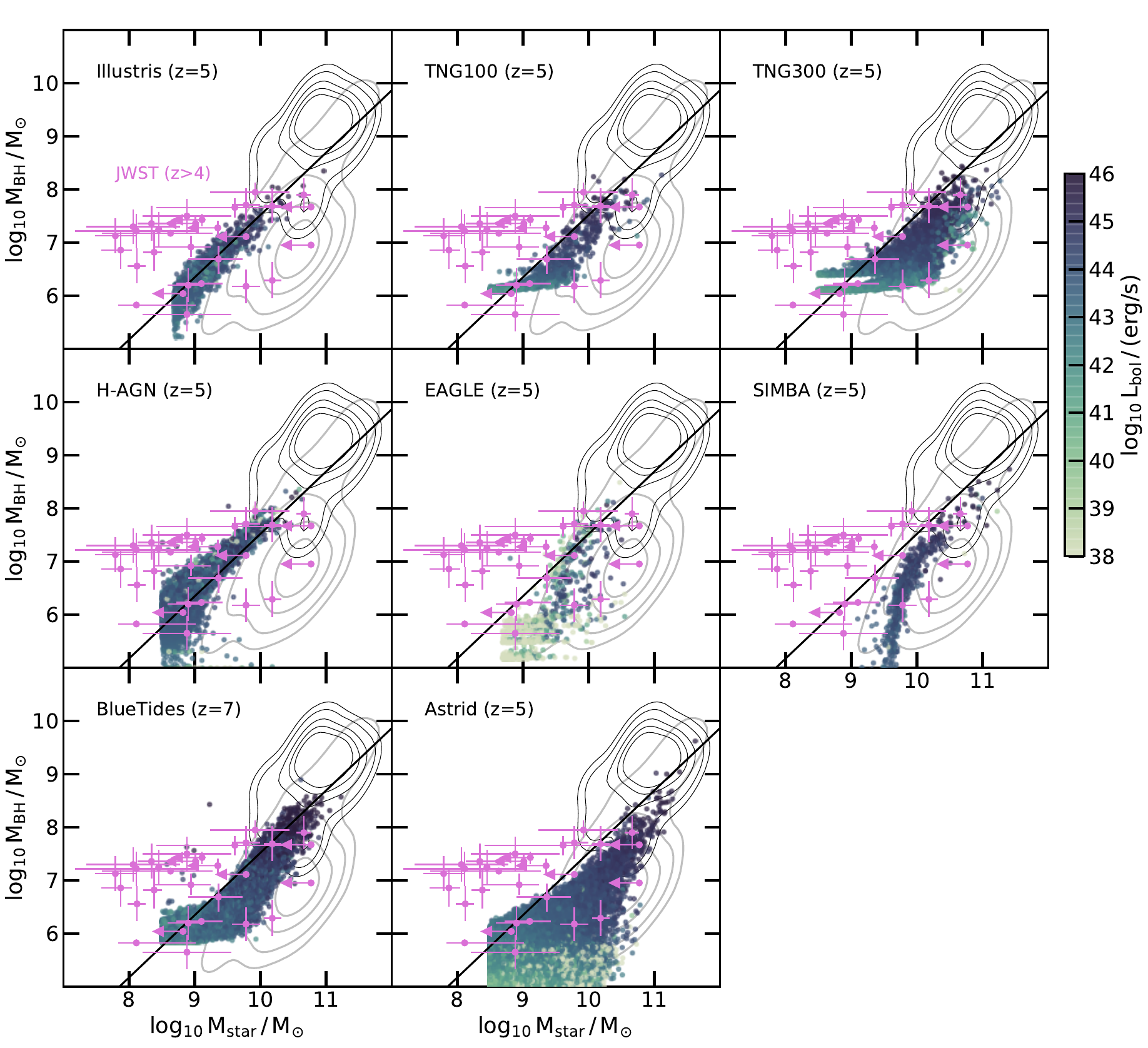}
    \caption{AGN candidates discovered at $z\geqslant 4$ with JWST \citep[purple data points,][]{2023ApJ...942L..17O,2023ApJ...946L..14K,2023arXiv230311946H,2023arXiv230512492M,2023arXiv230801230M,2023ApJ...951...72O} compared to current BH populations produced in large-scale cosmological simulations at $z=5$ ($z=7$ for BlueTides). We only considered simulated BHs located in resolved galaxies with $M_{\star}\geqslant 10^{8.5}\, \rm M_{\odot}$, and color code them by their bolometric luminosity (all BHs with $L_{\rm bol}\leqslant 10^{38}\, \rm erg/s$ are shown at that luminosity). 
    The sensitivity of JWST only allows for the detection of AGN with $L_{\rm bol}\geqslant 10^{44}\, \rm erg/s$, which should to be compared with the dark blue symbols representing the corresponding simulated AGN. To guide the eye, contours represent known high-redshift quasars prior JWST discoveries \citep[top contours][]{2019PASJ...71..111I}, the local population of BHs \citep[bottom contours][]{2015arXiv150806274R,2018ApJ...852..131B,2019arXiv191109678G}, and the empirical scaling relation from \citet{2013ARA&A..51..511K} is represented with a black line in each panel.}
    \label{fig:all_simulations}
\end{figure*}

We provide a summary of the BH population formed in the large-scale cosmological simulations at $z=5$ in Fig.~\ref{fig:all_simulations}. 
Simulations all produce different populations of BHs and AGN due to their different subgrid modeling \citep{2021MNRAS.503.1940H,2022MNRAS.509.3015H}. In Fig.~\ref{fig:all_simulations}, BHs are color coded by their AGN bolometric luminosity.
To guide the eye, we include in the figure the observational sample of \citet{2019PASJ...71..111I} for high-redshift quasars (assuming that $M_{\star}=M_{\rm dyn}$, upper contours) and observations in the local Universe \citep[][lower contours]{2015arXiv150806274R,2018ApJ...852..131B,2019arXiv191109678G}. 
The empirical scaling relation from \citet{2013ARA&A..51..511K} is represented with a black solid line in each panel.

Scaling relations, as we call them, were historically derived from samples of central BHs located in massive galaxies  \citep{Haring2004,MarconiHunt2003,McConnell2012a,2013ARA&A..51..511K}, often with $M_{\rm bulge}$ instead of the total stellar mass $M_{\star}$. As the only observations available at the time of the simulations' conception, they were used to calibrate them. As a result, their BH population tend to align well with scaling relations, particularly in the regime of massive galaxies.
More recent observational studies have revisited the $M_{\rm BH}-M_{\star}$ to account for the diversity of BHs observed in the local Universe \citep[e.g.,][]{2015arXiv150806274R,2015ApJ...798...54G}. Notably, incorporating BH mass estimates from 262 broad-line AGN and 79 inactive BHs with dynamical masses in galaxies with $M_{\star}=10^{9.5}-10^{11.5}\, \rm M_{\odot}$, as well as a few BHs located in dwarf galaxies of $\leqslant 10^{9.5}\, \rm M_{\odot}$, all at $z<0.055$, has led to a reevaluation of these relations, highlighting that using total stellar mass instead of bulge mass, which is often unavailable at high redshift, leads to less clear-cut and tight relations \citep{2015arXiv150806274R}.

As of today, the scatter in $M_{\rm BH}$ is of more than 3 dex at fixed $M_{\star}$ for $M_{\star}\geqslant 10^{10}\, \rm M_{\odot}$ in the local Universe. 
This scatter is not well reproduced in most simulations, mostly because of the lack of stochasticity in their subgrid models. The disparity is evident in Fig.~2 of \citet{2021MNRAS.503.1940H}, 
where simulated BHs at $z=0$ do not significantly overlap with the region occupied by observed low-redshift broad-line AGN (i.e., $M_{\rm BH}=10^{6}-10^{8}\, \rm M_{\odot}$ and $M_{\star}= 10^{10}-10^{11}\, \rm M_{\odot}$).
Astrid, the latest simulation, has a larger $M_{\rm BH}$ scatter, which is primarily due to a broader range of initial masses for the BHs and  
the longer timescales for low-mass BH coalescence as the simulation includes a model for the unresolved dynamical friction exerted by the surrounding collisionless matter (i.e., stars and dark matter in the Astrid's model). 
Some simulations also fail to produce BHs at fixed $M_{\star}$ as massive as observed (for example, for $M_{\star}\geqslant 10^{10}\, \rm M_{\odot}$).\\

We show the populations of simulated BHs at $z=5$ in Fig.~\ref{fig:all_simulations}, and compare them with the recent JWST high-redshift observations \citep[those available at the time of writting,][]{2023ApJ...942L..17O,2023ApJ...946L..14K,2023arXiv230311946H,2023arXiv230512492M,2023arXiv230801230M,2023ApJ...951...72O}. For simplicity, we report the candidates at $z=5$ although their actual redshift can go up to $z\sim 7-8$. We do not apply any selection biases to the simulated BHs and instead show the full populations produced in the simulations.
Simulated BHs align well with the AGN candidates located close to the local scaling relations. This is a consequence of the simulations' subgrid physics being tailored to reproduce these relations and BHs in most of these simulations evolve, indeed, from their formation in low-mass galaxies along the local scaling relations.
In contrast, most simulations fall short in reproducing the most massive AGN candidates in galaxies with $M_{\star}\leqslant 10^{10}\, \rm M_{\odot}$ discovered with JWST. This is simply because simulations were not calibrated to produce these overmassive BHs with high BH to stellar mass ratios.

\begin{figure*}
    \includegraphics[scale=0.46]{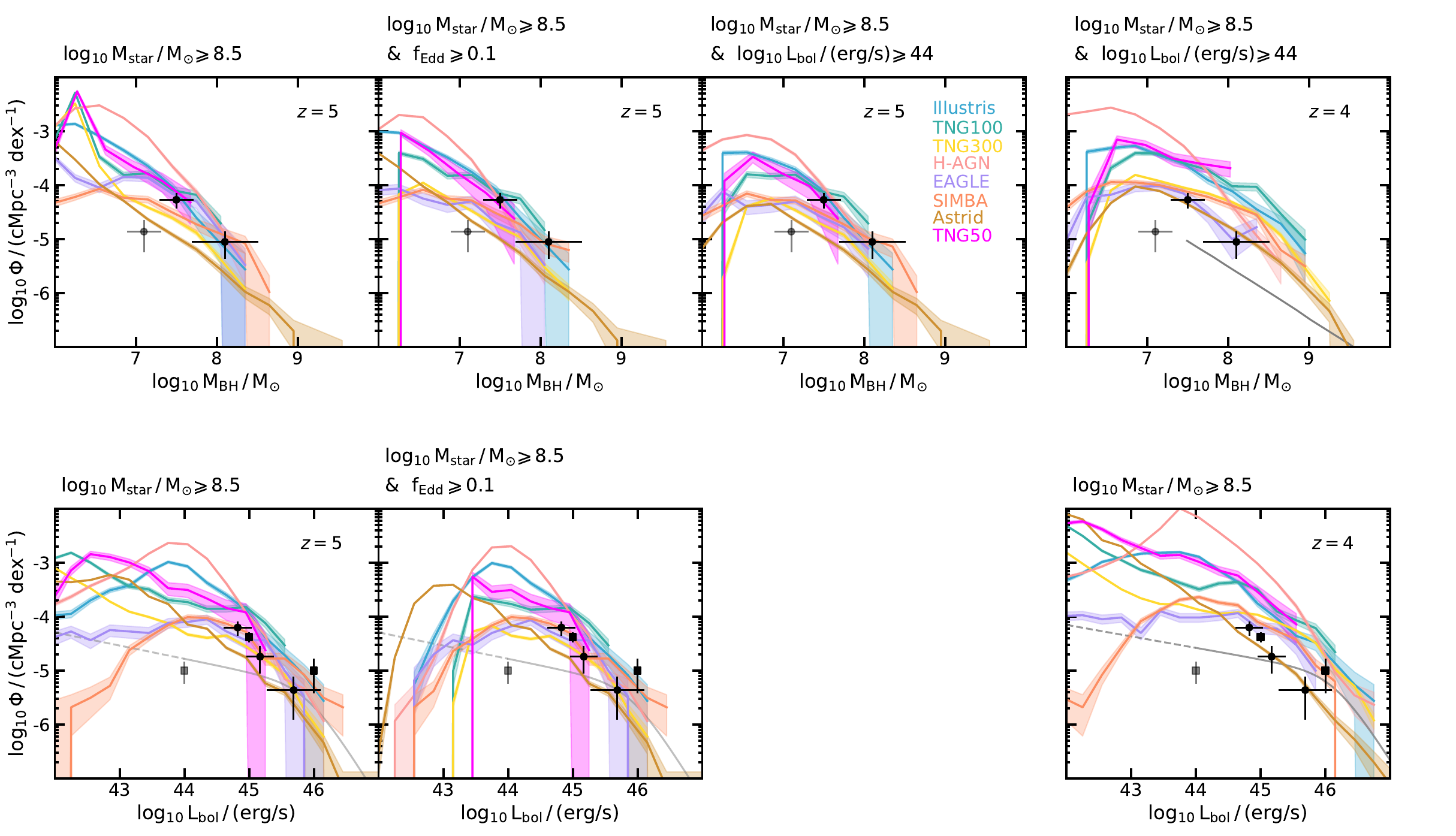}
    \caption{BH mass and bolometric luminosity function produced in the large-scale simulations of the field at $z=5$ ({\it 1st, 2nd, 3rd columns}) and $z=4$ ({\it 4th column}) for $M_{\star}\geqslant 10^{8.5}\, \rm M_{\odot}$. Shaded regions represent Poisson errorbars.
    Different cuts in BH mass, Eddington ratio, and bolometric luminosity are applied, as indicated in the title of the panels. 
    {\it Top panels:} Black circles represent observational constraints on the BH mass function at $z\sim 5$ (observations span $z=4-6$) from \citet{2023arXiv230605448M}. The first symbol at $\log_{10}\, M_{\rm BH}/\rm M_{\odot}\sim 7$ is shown in grey as it could suffer from incompleteness (see the above paper for details). The grey line on the right panel represents the observational constraints derived in \citet{2024ApJ...962..152H} from HSC-SSP/SDSS-DR7 and for broad-line AGN/quasars with $\log_{10}\, M_{\rm BH}/{\rm M_{\odot}}\geqslant 7.5$ at $z\sim4$.
    {\it Bottom panels:} Constraints on the H$_{\rm \alpha, \,broad}$ luminosity function from \citet{2023arXiv230605448M} that we convert to a bolometric luminosity function. 
    We add the constraints from \citet{2023arXiv230905714G} as black squares (observations span $z=4.5-6.5$); the first data points at $L_{\rm bol}\sim 10^{44}\, \rm erg/s$. 
    Grey lines represent the luminosity function derived in \citet{2020MNRAS.495.3252S} at $z=5$ or $z=4$ (for the two right panels, {\it 4th column}).}
    \label{fig:LF_all_simulations}
\end{figure*}

At high redshift, despite the remarkable advancements made possible by JWST observations, discoveries remain confined to a subset of the entire population of BHs theoretically predicted to exist. Observations primarily probe AGN powered by BHs with $M_{\rm BH}\geqslant 10^{6}\, \rm M_{\odot}$; evidence of AGN activity from lower-mass BHs can be found but disentangling them from other origins would remain highly unlikely, preventing any confirmation. As such, we cannot assess, from JWST observations alone, the scatter of the BH population at fixed $M_{\star}$. As most of the possible overmassive BHs are located in galaxies with $M_{\star}\leqslant 10^{8.5}-10^{9}\, \rm M_{\odot}$, this is also a difficult regime to compare with large-volume simulations which start resolving galaxies with enough particles for $M_{\star}\geqslant 10^{8.5}-10^{9}\, \rm M_{\odot}$.
In Fig.~\ref{fig:all_simulations}, the paucity of AGN detections with $M_{\rm BH}=10^{6-7}\, \rm M_{\odot}$ in galaxies with $M_{\star}\geqslant 10^{9.5-10.5}\, \rm M_{\odot}$ (i.e., where local AGN are typically observed) further raise the question of which BH population can be detected by JWST, as well as in which galaxies they have to be located to be detectable \citep{2017ApJ...849..155V}.
Consequently, the current JWST samples of AGN candidates may exhibit a strong bias toward active and massive BHs (likely located in relatively low-mass galaxies). Due to their lower-mass than observed with JWST, simulated BHs located in low-mass galaxies of $\leqslant 10^{9}\, \rm M_{\odot}$ have a hard time powering any AGN with $L_{\rm bol}\geqslant10^{44}\,\rm erg/s$ (the luminosities of the JWST AGN) in most cosmological simulations (see the color code of Fig.~\ref{fig:all_simulations}). Only simulated BHs in relatively massive galaxies with a characteristic mass of $M_{\star}\geqslant 10^{9.5}, \, 10^{10}\, \rm M_{\odot}$ (the exact value depends on the simulation) are able to grow efficiently and power these AGN. This happens when host galaxies are massive enough to overcome feedback from supernovae by retaining SN-driven winds avoiding gas removal from their galactic center.

Additionally, deriving accurate galaxy characteristics, such as stellar mass, age, metallicity, by SED fitting is hampered by several factors: degenerate parameters, simplified templates for the SFR history, uncertainties while removing the AGN contribution to the total system SED, for example. These effects can lead to uncertainties of $\sim1$ dex on $M_{\star}$. 
The JWST's NIRSpec/MSA prism enables the resolution of ionized gas kinematics, providing an independent constraint on galaxy velocity dispersion through 2D emission spectra and corresponding galaxy dynamical mass. This method was employed in \citet{2023arXiv230801230M} to assess whether possible outliers in the $M_{\rm BH}-M_{\star}$ diagram were also outliers in the $M_{\rm BH}-\sigma$ or $M_{\rm BH}-M_{\rm dyn}$ diagrams. Remarkably, AGN candidates identified as outliers in the JADES survey demonstrated broad agreement with the $M_{\rm BH}-\sigma$ and $M_{\rm BH}-M_{\rm dyn}$ relations observed in the local Universe. This highlights the complex interpretation of the current AGN candidates in the $M_{\rm BH}-M_{\star}$ plane.

For the reasons mentioned above, we refrain, in this paper, from drawing any conclusions from the JWST candidates regarding the potential relationship between BH mass and the stellar mass of their host galaxies at $z\geqslant 4$, nor its potential evolution with redshift \citep[but see][]{2024ApJ...964..154P}. In the large-scale simulations, there is no consensus on the time evolution of the mean $M_{\rm BH}-M_{\star}$ relation \citep{2022MNRAS.511.3751H}.
The overall normalization of the mean $M_{\rm BH}-M_{\star}$ relation decreases with decreasing redshift in, for example, Illustris, Horizon-AGN, and EAGLE. This trend can arise from a more efficient relative growth of galaxies compared to their central BHs at lower redshifts, likely due to less effective SN feedback compared to other simulations. In EAGLE, SN feedback impedes the initial BH growth in low-mass galaxies, while the BH's rapid growth phase occurs at a fixed halo virial temperature, indicating more substantial growth in more massive galaxies as redshift decreases.
Conversely, the normalization of the mean $M_{\rm BH}-M_{\star}$ relation increases with decreasing redshift in TNG100, TNG300, and SIMBA. This is due to higher BH growth rates at lower redshifts compared to their host galaxies, triggered by less effective SN feedback at low redshifts in TNG, while in SIMBA, it is primarily due to an increase in the galactic hot environment over time (attributed to AGN feedback), which in turn promotes an additional Bondi growth channel for the BHs.

\section{What about the abundance of AGN candidates?}
\label{sec:abundance}

To quantify the agreement between simulated and observed BHs, we show in Fig.~\ref{fig:LF_all_simulations} the BH mass function and the AGN bolometric luminosity function produced at $z=5$ and $z=4$ in the simulations. We only consider resolved galaxies with stellar mass of $\geqslant 10^{8.5}\, \rm M_{\odot}$. TNG50 is the only simulation in our sample which resolve lower-mass galaxies with $M_{\star}\geqslant 10^{7}\, \rm M_{\odot}$. However, the occupation fraction in galaxies with $10^{7}-10^{8.5}\, \rm M_{\odot}$ is below $30\%$ in TNG50 and the BHs located in these galaxies are inactive. The BH mass and AGN luminosity function are only very slightly affected by the addition of the low-mass galaxies, and as such we use the same galaxy mass threshold ($\geqslant10^{8.5}\, \rm M_{\odot}$) as for the other simulations for the figures.
This mass cut is not relevant for SIMBA which seeds BHs in galaxies of $M_{\star}\geqslant 10^{9.5}\, \rm M_{\odot}$; only those galaxies therefore appear in Fig.~\ref{fig:LF_all_simulations}. 
We compare these results to different observational constraints. 

First, we report on the top right panel the constraints on the AGN and quasar mass function derived in \citet{2024ApJ...962..152H} (their "DPL+LOG model", the grey line here) from HSC-SSP and SDSS-DR7 observations. Those are BHs with $\log_{10}\, M_{\rm BH}/{\rm M_{\odot}}\geqslant 7.5$ in the redshift range $z=3.5-4.25$.
On the bottom panel, we report the constraints on the bolometric luminosity function derived in \citet{2020MNRAS.495.3252S} from pre-JWST multi-band observations (the grey lines). The function is modeled as a double power-law: 
$\Phi=\Phi_{\star}/\left[ (L/L_{\star})^{\gamma_{\rm 1}}+(L/L_{\star})^{\gamma_{\rm 2}}\right]$,
with the following best-fit parameters (their ``local polished fit''): $\log_{10}\Phi_{\star}, \, \log_{10}L_{\star}, \, \gamma_{\rm 1}, \,  \gamma_{\rm 2}= -5.034, \,12.562,\, 0.213,\, 1.885$ at $z=4$, and $= -5.243, \,12.308,\, 0.245,\, 1.912$ at $z=5$.  The dashed part of the lines indicates the $L_{\rm bol}$ range that is not directly constrained but instead extrapolated from the observations used in the above paper.

Second, we show the recent JWST AGN samples of \citet{2023arXiv230605448M} from the FRESCO \citep{2023arXiv230402026O} and EIGER surveys and of \citet{2023arXiv230905714G} from the spectroscopic confirmation of the UNCOVER candidates \citep{2022arXiv221204026B,2023arXiv230607320L}. The more recent constraints of \citet{2024ApJ...968...38K} are also consistent with those. In the first study, AGN are selected with full width at half maximum of $\rm FWHM > 1000\, km/s$ and an H${\rm \alpha}$ luminosity of the broad component of $>2\times 10^{42}\, \rm erg/s$, and only selecting the systems with a well defined broad component ($\rm S/N_{H{\rm \alpha}}>5$).
The UNCOVER study selected well-defined red-color candidates with $\rm S/N_{F444W}>14$ and $\rm mag_{F444W}<27.7$ (within 0.32'' aperture), compact with a ratio between the flux measured with a 0.4 and 0.2'' apertures smaller than 1.7, and red in either F115W-F150W vs F200W-F356W and F200W-F444W or F150W-F200W vs F277W-F356W and F277W-F444W; broad-line AGN candidates are selected with $\rm FWHM > 2000\, km/s$ and same $\rm S/N_{H{\rm \alpha}}$ as above.
In both studies, BH masses are estimated from the \citet{2005ApJ...630..122G} relation calibrated at $z\sim 0$ and updated in \citet{2015arXiv150806274R}: $M_{\rm BH}\propto (L_{\rm H{\alpha}})^{\gamma}\times  ({\rm FWHM}_{\rm \alpha})^{\gamma}$.

We directly employ the constraints on the bolometric luminosity function from \citet{2023arXiv230905714G}. However, we translate the H${\rm\alpha,broad}$ constraints of \citet{2023arXiv230605448M} into constraints on bolometric luminosities (which likely induce an enlargement of the uncertainties) with the following relation derived from \citet{Richards06}:
$L_{\rm bol}= 10.33 \times (L_{\rm H{\alpha,\, broad}}/5.25/10^{42})^{1/1.157} \times 10^{44}\, \rm erg/s$.

Finally, the FRESCO and EIGER observations span from $z=4.2$ to $z=5.5$, but are centered at $z\sim 5$. The surveys cover a volume of $5.7\times 10^{5}\, \rm cMpc^{3}$. 
The volume of the UNCOVER survey is more delicate to compute because lensing distortion of the galaxy cluster Abell 2744 field \citep[][]{2023arXiv230905714G}. The UNCOVER AGN are located at redshift $z=5$ to $8.5$, but we confine the analysis to the constraints derived in $z=5-6$. Other samples exist in the literature, such as JADES \citep{2023arXiv230512492M}, but we restrain from using them due to the non-trivial estimation of the volume they cover with their spectroscopic selections.
In comparison to the surveys mentioned above, TNG50 has a volume of $10^{5}\, \rm cMpc^{3}$, while most of the other simulations shown in Fig.~\ref{fig:LF_all_simulations} simulate a volume that range from $10^6\,\rm cMpc^{3}$ (Illustris, TNG100, EAGLE, SIMBA, and a slightly larger volume for Horizon-AGN) to $\sim 10^{7}\,\rm cMpc^{3}$ (TNG300, Astrid). The above comparison of the observed and simulated volumes reassures us that the comparisons of the BH mass or AGN luminosity functions are meaningful.

\subsection{BH mass function}
The discrepancies in the BH mass function from the different simulations is within 1 dex at $\log_{10}\, M_{\rm BH}/{\rm M_{\odot}}\geqslant 8$, and increases to more than 1 dex towards lower-mass BHs. When considering all simulated BHs independently of their Eddington ratios or bolometric luminosity ({\it top left panel} of Fig.~\ref{fig:LF_all_simulations}), the effect of the seeding, accretion, SN feedback is directly visible on the mass function. For example, the peak of the TNG and EAGLE BH mass functions at $\log_{10}\, M_{\rm BH}/{\rm M_{\odot}}\leqslant 6$ represents the seed population, and the subsequent decrease of the functions at slightly higher BH mass represents the ungrown BHs in these simulations with an inefficient BH growth in low-mass galaxies (due to the lack of efficient accretion or efficient SN feedback, or a combination of these effects). As driven by inactive BHs, these effects in the $\log_{10}\, M_{\rm BH}/{\rm M_{\odot}}=6-7$ mass range cannot be constrained by JWST. They, indeed, mostly vanish when considering only the detectable active BHs ({\it middle panels}, with $f_{\rm Edd}\geqslant 0.1$ or $L_{\rm bol}\geqslant 10^{44}\, \rm erg/s$). In the observations, the mass range of $\sim 10^{7}\, \rm M_{\odot}$ can be probed by JWST for some active BHs but suffers the most from incompleteness in the observational samples due to the H${\rm \alpha}$ luminosity threshold used in their selections: only the most accreting BHs in this mass range would power bright enough AGN to be detected. However, constraining the active BH mass function in this mass range would provide us with key constraints to disentangle a combination of the subgrid models employed in simulations.

Regarding more massive BHs of $\log_{10}\, M_{\rm BH}/{\rm M_{\odot}}\geqslant 7.5$, the constraints on the BH mass function of \citet{2023arXiv230605448M} appear to have a higher normalization than the function of Astrid and TNG300, and a similar normalization as TNG50, TNG100, Illustris, and EAGLE ({\it top left panel} of Fig.~\ref{fig:LF_all_simulations}, still considering all active and inactive BHs). The difference for the TNGs arises from the different resolutions: TNG300 has the lowest resolution, which leads to an overall BH population that accrete gas less efficiently and do not grow as much as in TNG100 or TNG50.
We note that Horizon-AGN produces more BHs of $\log_{10}\, M_{\rm BH}/{\rm M_{\odot}}\sim 7.5$ and fewer BHs of $\sim 8$ compared to the observations; SIMBA produces the opposite trend.

In the observations, all AGN candidates are estimated to have bolometric luminosities above $10^{44}\, \rm erg/s$. Therefore, we operate a cut of $f_{\rm Edd}\geqslant 0.1$ and $\geqslant 10^{44}\, \rm erg/s$ in the {\it 2nd and 3rd-column panels} and as such only consider the active BH mass function. These cuts decrease the number density of simulated BHs and tensions start arising when compared to the observational constraints: only one or a couple of simulations remain above the constraints at a given $M_{\rm BH}$. 

These puzzling results are enhanced by the fact that our BH mass functions for the simulations assume that all BHs are ``visible'' (i.e., a null obscuration fraction). To the contrary, however, it would not be suprising if the broad-line AGN detected by JWST were only a subpopulation of the entire BH population at $z\sim 5$. The fraction of obscured AGN with $L_{\rm bol}\sim 10^{44}\, \rm erg/s$ and column densities of $\log_{10}\, \rm N_{H}/\, cm^{-2} \sim 22$ could be as high as $80\%$ at $z\geqslant 5$ \citep[][based on samples of massive galaxies with $M_{\star}> 10^{10}\, \rm M_{\odot}$]{2022A&A...666A..17G}.\\

For comparison, we show the BH mass function produced in the simulations at $z=4$, alongside the same $z=5$ JWST observational constraints. We also show the constraints from \citet{2024ApJ...962..152H} at $z\sim 4$. Interestingly, JWST constraints lie about 0.5 to 1 dex above the broad-line AGN/quasars from HSC-SSP/SDSS-DR7 ones.
The more evolved simulated BH populations at $z\sim 4$ are in better agreement with the JWST constraints than the simulated BHs at $z\sim 5$, but a few simulations (e.g., EAGLE and Astrid) still fall short in producing a population larger that considering the observed candidates alone. This is very interesting because the most recent Astrid simulation, for example, has a larger $M_{\rm BH}$ scatter at fixed $M_{\star}$ and as a result a good agreement between the simulated BH population at $z=3$ (last snapshot of Astrid) and the observed $M_{\rm BH}-M_{\star}$ diagram observed in the local Universe \citep{2022MNRAS.513..670N}. If the constraints from JWST are confirmed with more observational samples and smaller uncertainties on BH mass estimates, we will need to re-assess the assembly of BHs in simulations and scrutinize the possible causes of differences in the high and low-redshift Universe.

\begin{figure*}    
    \includegraphics[scale=0.46]{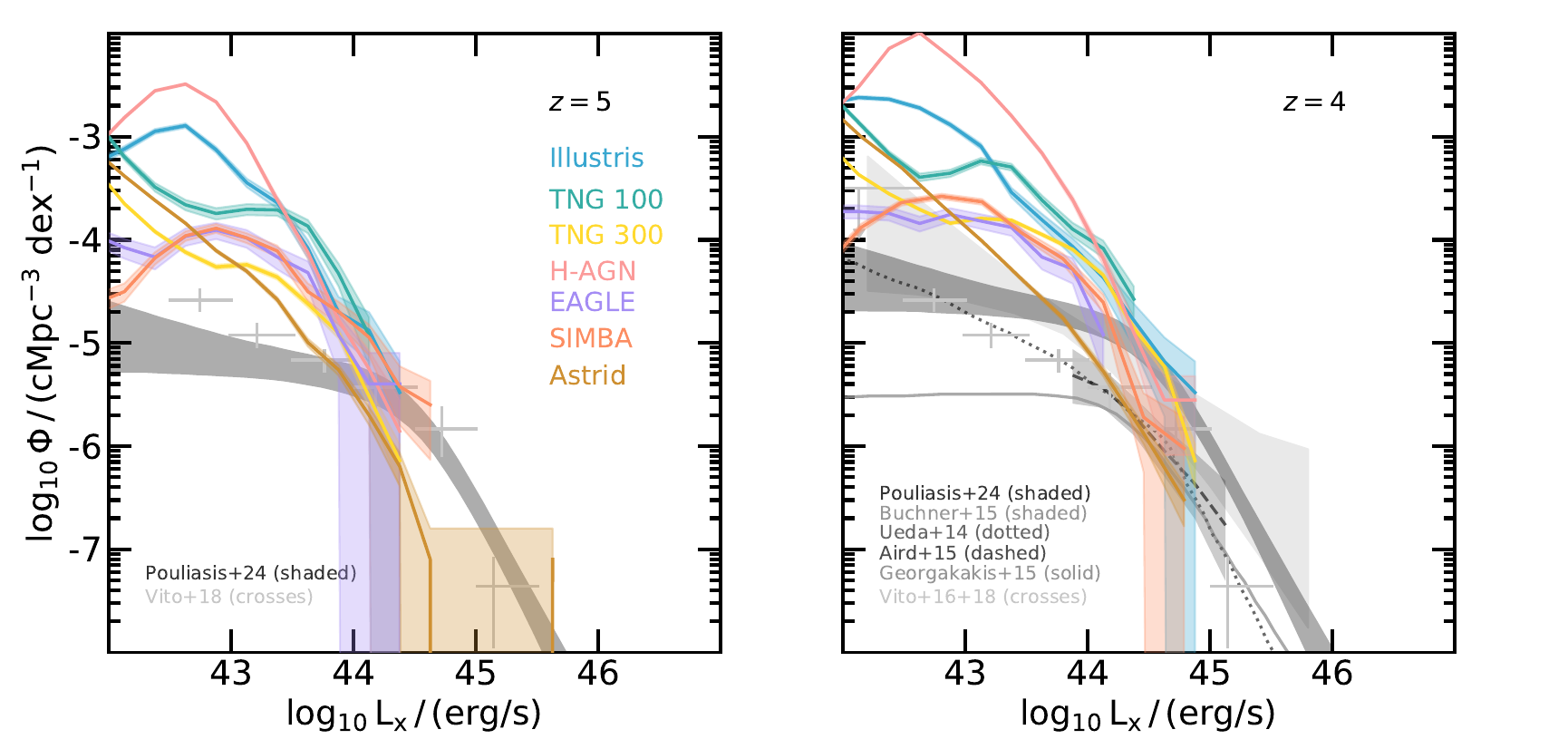}
    \caption{X-ray (2-10 keV) luminosity function of the simulations of the field at $z=5$ and $z=4$ for $M_{\star}\geqslant 10^{8.5}\, \rm M_{\odot}$, compared to the observational constraints derived in \citep{2014ApJ...786..104U,2015MNRAS.451.1892A,2015MNRAS.453.1946G,2015ApJ...802...89B,2016MNRAS.463..348V,2018MNRAS.473.2378V,2024arXiv240113515P}.
    The most recent constraints from \citet{2024arXiv240113515P} are integrated over the redshift range $z=4.5-5.5$ and $z=3.5-4.5$, and $\log_{10} N_{\rm H}/\rm cm^{-2}=(20,26)$. Constraints from \citet{2016MNRAS.463..348V} are for the range $z=3.6-6$ (reported on left and right panels). 
    See text for the description of the other constraints shown in the $z=4$ panel.
    Overall, when assuming the X-ray bolometric correction of \citet{2020MNRAS.495.3252S} most of the X-ray luminosity functions produced in the simulations agree fairly well with the bright end of the observational functions ($L_{\rm 2-10\, kev}\sim10^{44}\, \rm erg/s$ at $z=5$), and overestimate their faint end.
    The agreement at the bright end improves when adding a correction for unresolved short timescale AGN variability (not shown here) or the bolometric correction of \citet{2020A&A...636A..73D}. }
    \label{fig:XrayLF_all_simulations}
\end{figure*}

\subsection{AGN bolometric luminosity function}

We show in the {\it bottom right panels} of Fig.~\ref{fig:LF_all_simulations} the comparison for the bolometric luminosity function. The regime covering the observational constraints is driven by BHs of at least $\geqslant 10^{6}\, \rm M_{\odot}$ and with $f_{\rm Edd}\geqslant 0.1$, and is therefore not affected by the cut in the Eddington ratio that we show in the {\it 2nd-column panel}.

First, the slope of the bright-end of most of the simulations agrees fairly well with the observational constraints of \citet{2023arXiv230605448M,2023arXiv230905714G}. The bright end of the luminosity function is in correct agreement with the pre-JWST constraints of \citet[][grey solid line]{2020MNRAS.495.3252S}, while the simulations all produce more faint AGN with $L_{\rm bol} \leqslant 10^{45}\, \rm erg/s$.
For some simulations, the knee of the function (i.e., the most common AGN in galaxies with $\geqslant 10^{8.5}\, \rm M_{\odot}$) appear at slightly fainter AGN ($\sim 10^{44}\, \rm erg/s$) than the ones detected with JWST (e.g., Horizon-AGN, Illustris, EAGLE). However, TNG50, TNG100, and Astrid have broader $L_{\rm bol}$ distributions or peak at smaller $L_{\rm bol}$. In Astrid, this is because the simulation produces a larger scatter in $M_{\rm BH}$, mostly driven by the use of a power-law distribution to initiate BH seed masses ($M_{\rm seed}=3\times 10^{4}- 3\times 10^{5}\, h^{-1}\, \rm M_{\odot}$) instead of a fixed initial mass. The high number of AGN with $L_{\rm bol}\sim 10^{42-43}\, \rm erg/s$ in the TNGs represent seed-mass BHs ($\sim 10^{6}\,\rm M_{\odot}$) prevented from accreting at their Eddington rate because of an efficient feedback from SNe at high redshift (these systems disappear when applying a $f_{\rm Edd}\geqslant 0.1$ cut, {\it 2nd-column panel}). The cut in $f_{\rm Edd}$ does not impact the $L_{\rm bol}$ regime probed by JWST.
According to the bolometric luminosity function of the simulations, JWST could be either detecting a small subset of the high-redshift AGN population (the most efficiently growing that are far from the bulk of the population, this is the case for the TNG simulations for example) or is actually detecting a subset of AGN growing only slightly more efficiently than the bulk of the population (e.g., SIMBA). 

When focusing on AGN that are most likely detected with JWST, those with $f_{\rm Edd}\geqslant 0.1$ and $L_{\rm bol}\geqslant 10^{44.5}\,\rm erg/s$, we find that several simulations produce a luminosity function matching the constraints, with some even falling below the constraints (e.g., Astrid). This, again, comes as a surprise as we do not apply any correction for obscuration.
Conversely, certain simulations (e.g., Horizon-AGN, TNGs, Illustris) predict a higher abundance of these relatively bright AGN, which alleviates the tension between simulations and observed luminosity functions. These simulations have a population of BHs evolving along the local scaling relations at high redshift either for the full galaxy stellar mass range ($M_{\star}\geqslant 10^{8.5}\, \rm M_{\odot}$, such as in Horizon-AGN and Illustris) or when galaxies are massive enough ($M_{\star}\geqslant 10^{9.5}\, \rm M_{\odot}$) to overcome SN feedback such as in TNG (see the shape of the $M_{\rm BH}-M_{\star}$ diagram in Fig.~\ref{fig:all_simulations}). 

These simulations illustrate the degenerate effect of seeding and growth of the BHs to produce an abundance of massive BHs with $L_{\rm bol}\geqslant 10^{44.5}\, \rm erg/s$: while a high seed mass is needed in TNG ($M_{\rm seed}\sim 10^{6}\, \rm M_{\odot}$) to balance the efficient regulation of BH growth by SN feedback in high-redshift low-mass galaxies, the more efficient early growth of the BHs in Horizon-AGN and Illustris eases the requirement on BH initial mass \citep[see also][]{2022MNRAS.514.4912H}. We note that in Horizon-AGN the seeding is based on prescriptions on the local gas properties: BHs form in dense gas cells (i.e., in cells for which the gas density exceeds the threshold for star formation) with stellar velocity dispersion greater than 100 km/s \citep{2016MNRAS.463.3948D}.
As a result, BH seeds tend to form earlier (in lower-mass galaxies) than in some of the other simulations. This yield to BHs of $10^{6}\, \rm M_{\odot}$ in galaxies with $10^{8.5}\, \rm M_{\odot}$ in Horizon-AGN, similar to the TNG BHs, although the seed mass in Horizon-AGN is of $\sim 10^{5}\, \rm M_{\odot}$ instead of $\sim 10^{6}\, \rm M_{\odot}$ for TNG.

In the {\it left bottom panel} in Fig.~\ref{fig:LF_all_simulations}, we show a comparison between the more evolved simulated populations at $z=4$ and the observational constraints at $z\sim 5$, which again provides a better agreement.\\

For context, most of the simulations studied here agree relatively well with the constraints coming primarily from X-ray observations at $z\sim 5$ and $z\sim 4$, i.e., with the latest constraints derived in \citet{2024arXiv240113515P} and previous works \citep[e.g.,][]{2014ApJ...786..104U,2015MNRAS.451.1892A,2015MNRAS.453.1946G,2015ApJ...802...89B,2016MNRAS.463..348V,2018MNRAS.473.2378V}. 
In Fig.~\ref{fig:XrayLF_all_simulations}, we show the parametric function 
from \citet{2024arXiv240113515P}, integrated over the range $\log_{10} N_{\rm H}/\rm cm^{-2}=(20,26)$ and $z=4.5-5.5$ and $z=3.5-4.5$ (their Fig.~10, but including here the Compton-thick regime, private comm. with Ektoras Pouliasis). 
The constraints on the total luminosity function for $z\sim 5$ do not exceed $\sim 6\times 10^{-6}$ at $L_{\rm x}=10^{44}\, \rm erg/s$. 
The constraints agree well with those presented in \citet{2018MNRAS.473.2378V}, which cover the range $z=3.6-6$ and that we reproduce in our two panels. 
The constraints from \citet{2015MNRAS.451.1892A} that we show on our $z=4$ panel are for the full range of column densities ($\log_{10} N_{\rm H}/\rm cm^{-2}=(20,26)$), include unobscured, Compton-thin, and -thick sources, and cover the redshift range $z=3.5-5$; we only show the fit from the hard-X-ray band sample (their Fig.~8). This study finds a total luminosity function  
not exceeding a few $10^{-5}$ and $10^{-6}$ for $L_{\rm x}=10^{44}$ and $10^{45}\, \rm erg/s$, respectively, at $z=3$ (their Fig.~9). Fig.~\ref{fig:XrayLF_all_simulations}. 
Because of the redshift evolution of the luminosity function found in observational studies, 
the normalization of the constraints from these two studies could be lower for $z=5$ than shown in Fig.~\ref{fig:XrayLF_all_simulations}.
On the $z=4$ panel, the results from \citet{2015ApJ...802...89B} are for the total luminosity function and the range $z=4-7$,  constraints from \citet{2014ApJ...786..104U} are only corrected for Compton-thin AGN and are for $z=4-5$ \citep[once corrected for Compton-thick, results are similar to those of][]{2024arXiv240113515P}. Because integrated from $z=4$ to higher redshifts, these constraints may represent lower limits at $z=4$.
Finally, the function of \citet{2015A&A...578A..83G} is for $z\sim 4.1$ and does not include a correction for X-ray obscuration in the determination of the AGN X-ray luminosities.
Differences in the observed constraints depend on many different aspects, primarily corrections for obscuration (Compton-thin, Compton-thick sources) but also, for example, how redshift uncertainties are considered.

To derive the X-ray luminosities of the simulated AGN, we employ the X-ray bolometric correction of \citet{2020MNRAS.495.3252S}:
$L_{\rm bol}/L_{\rm 2-10\, keV} = \alpha_{1} (L_{\rm bol}/{\rm 10^{10}\, L_{\odot}})^{\beta_{1}} + \alpha_{2} (L_{\rm bol}/{\rm 10^{10}\, L_{\odot}})^{\beta_{2}}$,
with $\alpha_{1}, \alpha_{2}=12.60, 4.073$ and $\beta_{1}, \beta_{2}=0.278, -0.026$. 
In light of the recent JWST results \citep[and the possible non-detection of the JWST broad-line AGN candidates in X-ray,][]{2024arXiv240413290Y,2024arXiv240419010A,2024arXiv240500504M}, this approach that has been widely used so far in the literature may be too simplistic but we prefer it to the computational expensive modeling of the SED of all the simulated galaxies and AGN of all these simulations. As observational constraints are already corrected for incompleteness, absorption, obscured and unobscured sources, we do not correct the luminosity function of the simulations further.

Overall, simulations produce more abundant AGN with $L_{\rm 2-10\, keV}\leqslant 10^{44}\rm \, erg/s$ at $z\sim 5,\,4$ with respect to the observational constraints \citep[especially for recent studies constraining the function up to $z\sim 6$,][]{2024arXiv240113515P,2018MNRAS.473.2378V}. 
When available for the observational samples, constraints on the host's stellar mass indicate that the X-ray AGN are hosted in relatively massive galaxies with $M_{\star}\geqslant 10^{10}\, \rm M_{\odot}$. A large fraction of faint AGN are produced in galaxies with $<10^{10}\, \rm M_{\odot}$ in the simulations, and only keeping more massive galaxies reduces significantly the faint end of the luminosity functions and yield a better agreement with observations \footnote{A similar trend is obtained when applying a correction for obscured AGN to the simulations. This is the case when assuming for example the constraints on the obscured fraction from \citet{2014ApJ...786..104U, 2014MNRAS.437.3550M} with about $80\%$ of AGN with $L_{\rm x}\sim 10^{44}\, \rm erg/s$ being obscured and $<40\%$ for $L_{\rm x}\geqslant 10^{45}\, \rm erg/s$.}. This indicates that the subgrid physics of simulations either produce too many faint AGN in galaxies with $<10^{10}\, \rm M_{\odot}$ or that current surveys in X-ray do not capture those \citep[see discussions in][]{2022MNRAS.509.3015H,2024arXiv240113515P}.
This aspect requires to be explored further as JWST AGN candidates with $M_{\star}$ estimates seem to be primarily hosted in such low-mass galaxies, with first analyses of Chandra stacking of broad-line candidates not showing a strong X-ray signal.
In any case, the X-ray luminosity range that would correspond to $L_{\rm bol}\sim 10^{44.5}\rm \, erg/s$ (i.e., the lower end probed by JWST that does not significantly suffer from incompleteness) seems to provide a better agreement between simulations and observational constraints.

For brighter AGN with $L_{\rm 2-10\, keV}\sim 10^{44}\, \rm erg/s$, some simulations also produce more of those or instead produce similar number densities agreeing with the lower-hand of the constraints at $z\sim 4$ \citep[e.g., Astrid, see also][]{2022MNRAS.513..670N}. The conclusions that can be drawn from the comparison of the X-ray 2-10 keV luminosity functions at $z\sim 5$ are less strong but resemble those we made for the bolometric luminosity function, keeping in mind that there is a good general agreement at $z\sim 4$ where observational X-ray data are more abundant.
The agreement at the bright end (mostly for $L_{\rm 2-10\, keV}\geqslant 10^{44}\rm \, erg/s$) improves when adding a correction for unresolved short timescale AGN variability \citep[not shown in Fig.~\ref{fig:XrayLF_all_simulations}, but demonstrated in][]{2022MNRAS.509.3015H}. The luminosity functions also slightly shift towards brighter AGN when employing the bolometric correction derived in \citet{2020A&A...636A..73D}.

\subsection{Interpretation of the comparison between observed and simulated AGN abundances}
In general we note a stronger agreement between the simulations and the constraints on the luminosity function compared to the mass function. Below, we outline potential avenues to elucidate the puzzle that the first observational constraints from JWST have triggered.
Two options are presented here, depending on whether one aims to draw conclusions from an observational standpoint or from the perspective of simulations.

First, uncertainties may arise from the challenges inherent in JWST observations and the complexities involved in estimating AGN characteristics:
\begin{itemize}
    \item Not all the AGN candidates may actually be AGN, or they may not have the currently estimated properties. Broad-line AGN are primary selected based on the broadening of their H$\alpha$ emission line, with large FWHM of at least 2000 km/s. Other processes could produce this broadening, or contribute to it. For example, this could be the case for outflows generated by AGN at larger scales than their broad-line region, outflows driven by star formation or SNe, and contamination from a close star-forming gas clump or star formation from a neighboring system. While these processes can produce lines as broad as those observed by JWST (i.e., FWHM>1000-2000 km/s), there is a consensus that outflows or contaminants typically produce additional signatures that aid in distinguishing them from AGN, such as a noticeable shift between the broad and narrow components and/or the presence of other broad lines (e.g., broad component of the OIII doublets). 
    \item The mass and/or luminosity of observed AGN could be overestimated. This could be caused by intrinsic uncertainties of the virial relations derived in the local Universe, potential evolution of these relations with redshift, different geometries of the broad-line region \citep[e.g., its size,][]{2008ApJ...678..693M,2016ApJ...825..126D,2024A&A...684A.167G}, or the other assumptions made (e.g., broad-line region not in equilibrium, BHs in super-Eddington regime), an unaccounted-for large contribution from outflows or other additional components in the broadening of the observed H${\alpha}$ line. The first aspects are currently being discussed as results on X-ray stacking of the broad-line AGN candidates are emerging and showing no X-ray counterparts to the JWST observations for a significant fraction of the AGN \citep{2024arXiv240413290Y,2024arXiv240419010A,2024arXiv240500504M}, although intrinsic X-ray weakness cannot be excluded \citep[][for a discussion on the implications for optical to X-ray spectral indices]{2024arXiv240214706I}.
    \item As pointed out in \citet{2023arXiv230905714G}, there are several factors which enter into the luminosity function that are difficult to compute (i.e., the volume covered in magnified fields, the completeness of the samples for the bolometric luminosity function).
\end{itemize}

Conversely, if we assume that all the observed red broad-line AGN are indeed AGN and that their derived bolometric luminosity and BH masses are accurate, we can draw the following conclusions regarding the simulations studied here and their subgrid modeling:
\begin{itemize}
    \item Since we did not include any fraction of obscured AGN for the simulations, if the observed abundance of AGN is accurate it suggests that, according to the simulations studied here, JWST is currently detecting a significant portion of the full predicted population of AGN with $L_{\rm bol}=10^{44.5}-10^{46}\, \rm erg/s$ (through their broad H$\alpha$).
    \item The simulations whose AGN number density underestimates or matches the constraints do not produce enough luminous or luminous enough AGN, and possibly do not produce enough massive BHs to power those. Several effects could (to some extent) help reducing the discrepancies with observations. With the same BH masses, brighter AGN can be obtained by including post-processing the unresolved short timescale variability of the AGN; this results in an enhancement of the bright end of the luminosity function. 
    \item Faster assembly of massive BHs could be driven by the use of a BH initial mass distribution skewed towards massive seeds, more efficient accretion, or less efficient SN and/or AGN feedback in low-mass galaxies \citep[][for a study on the relative roles of these feedback processes]{2022MNRAS.516.2112K}, or a combination of these effects. These aspects would help reconcile simulated BHs with the high $M_{\rm BH}/M_{\star}$ observed AGN in the low-mass galaxy regime of the  $M_{\rm BH}-M_{\star}$ diagram. 
    \item These simulations do not model super-Eddington accretion \citep[see][for a recent implementation]{2023A&A...670A.180M} but instead cap $\dot{M}_{\rm acc}$ to the Eddington limit $\dot{M}_{\rm Edd}$. In the super-Eddington regime, BHs can accrete above that limit while their luminosity scales as $\log_{10}\dot{M}_{\rm acc}$ with a reduced radiative efficiency so that more mass is accreted into the BHs; their luminosity, however, does not significantly exceed their Eddington luminosity. Although these episodes would favor the fast assembly of massive BHs \citep{2023arXiv230512504S}, their feedback could stunt their growth after only a few Myr, as shown in high-resolution hydrodynamical simulations \citep{2023A&A...670A.180M}. 
    
    \item Modifications of the simulation subgrid physics in the above-mentioned directions may require however a redshift dependence of the processes to avoid worsening the agreement between the populations of simulated BHs and observations in the local Universe. For example, BHs with high $M_{\rm BH}/M_{\star}$ ratios in low-redshift dwarf galaxies are generally unobserved at $z\sim0-2$ \citep[but see][]{2023ApJ...943L...5M,2024arXiv240405793M}. Conversely, faster growth of BHs at high redshift could help producing the most massive BHs observed at fixed galaxy stellar mass in the local Universe, for $M_{\star}\geqslant 10^{10}\, \rm M_{\odot}$, that are currently not well reproduce in the simulations \citep[see][for a detailed study]{2020arXiv200610094H}. 
    \item We emphasize here that the BH occupation fraction of galaxies with $M_{\star}\geqslant 10^{8.5}\, \rm M_{\odot}$ almost reaches unity at $z\sim 4-5$ in all the simulations studied here. A large occupation fraction of BHs in lower-mass galaxies ($<10^{8.5}\, \rm M_{\odot}$) would only boost the bright end of the AGN luminosity function if these BHs have $M_{\rm BH}/M_{\star}$ in between 0.1 to unity, while increasing the normalization of the AGN mass function. This could reconcile the $M_{\rm BH}-M_{\star}$ diagram with high $M_{\rm BH}/M_{\star}$ systems and the AGN mass and luminosity functions.
    \item Due to limited resolution, large-scale simulations cannot self-consistently model the coalescence of BHs and only a few simulations have a subgrid model to account for the first ``dynamical friction'' phase of the mergers \citep[Horizon-AGN and Astrid,][]{2020MNRAS.498.2219V,2022MNRAS.510..531C,2022MNRAS.513..670N}. A common model in the other simulations is to merge BHs when their host galaxies merge, which favors the growth of BHs (even if BH mergers only represent a secondary channel for BH growth, behind BH accretion). More accurate modeling of BH dynamics would not foster BH growth. This in turn also increases growth by accretion, as more massive BHs are allowed to accrete more in the simulations: Bondi-Hoyle-Lyttleton accretion scales as $M_{\rm BH}^2$ and the Eddington cap scales as $M_{\rm BH}$. Therefore if BHs are merged `prematurely’ they will also see their accretion boosted \citep[see Appendix in][]{2020MNRAS.498.2219V}.

    \item  The closer agreement between the observational constraints at $z\sim 5$ and certain simulated luminosity functions at $z=4$ suggests that in some simulations, the overall growth and assembly of simulated BHs at high redshift could be slower and/or delayed by 300-400 Myr.

    \item Finally, the brighter AGN that would be required to agree with the observational constraints need a sustained gas reservoir in the host galaxies, i.e. a quick replenishment of cold gas, to sustain the efficient growth of the BHs. Conversely, to not exceed the current number density of more massive BHs in the quasar regime \citep[at the moment, JWST AGN candidates are about 100 times more abundant than faint quasars, e.g.,][]{2018ApJ...869..150M} these AGN will need to be regulated at some point (by quenching themselves or through the processes mentioned above such as SN feedback, starvation, BHs leaving the central gas reservoir). 

\end{itemize}

\begin{figure*}    
    \includegraphics[scale=0.55]{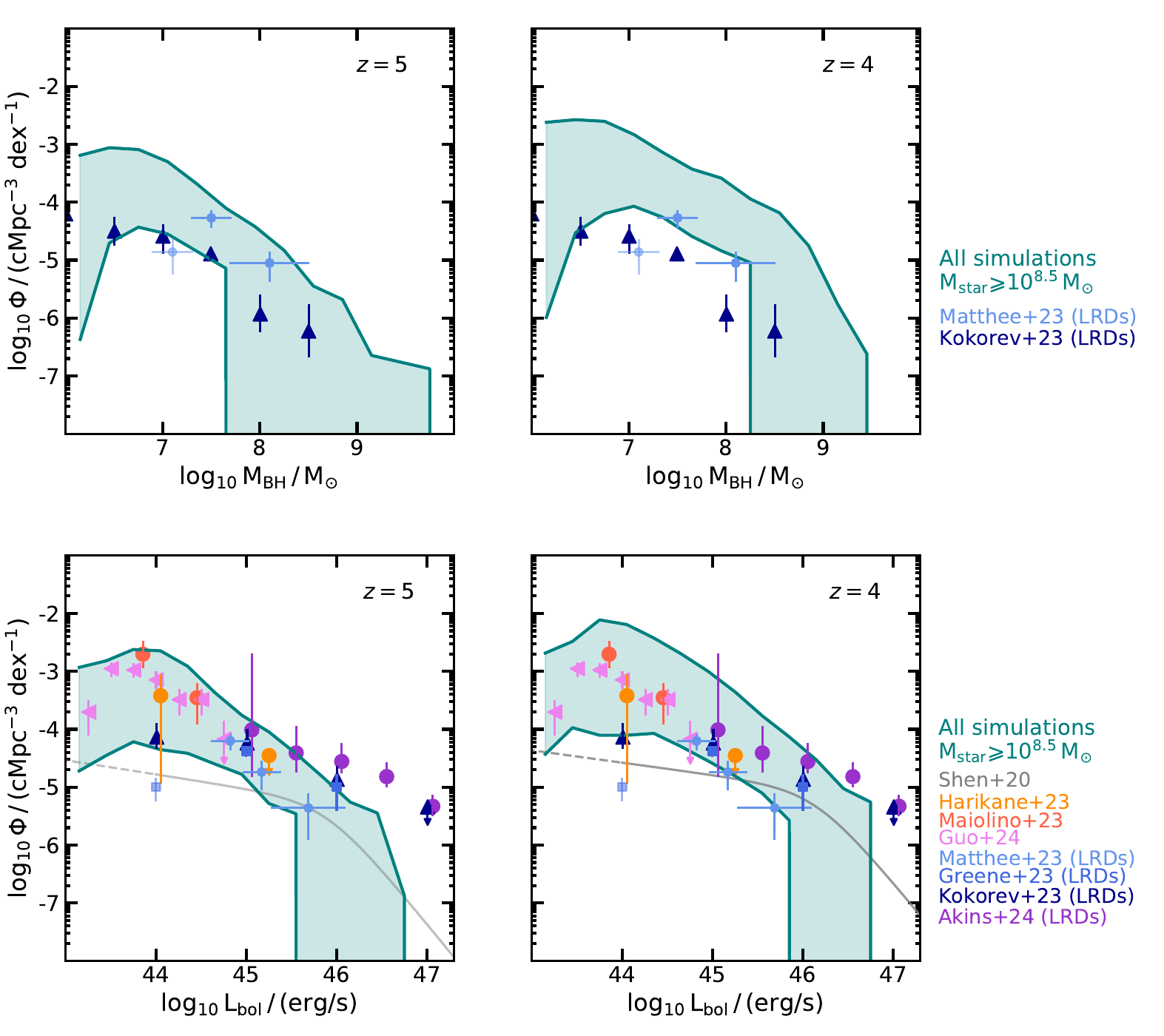}
    \caption{Summary of the BH mass and bolometric luminosity functions for the large-scale simulations of the field at $z=5$ ({\it left panels}) and $z=4$ ({\it right panels}), and for $M_{\star}\geqslant 10^{8.5}\, \rm M_{\odot}$. For the BH mass function, we include all BHs with $L_{\rm bol}\geqslant 10^{44}\, \rm erg/s$.
    Symbols represent observational constraints from JWST, from sample of AGN candidates or candidates selected as LRDs observed at $z\sim 5$ with some spanning the $z=4-6.5$ range \citep{2023arXiv230605448M,2023arXiv230905714G,2023arXiv230801230M,2023arXiv230311946H,2024arXiv240610341A,2024arXiv240919205G,2024ApJ...968...38K}.
    Grey lines represent the luminosity function derived from multi-band pre-JWST observations in \citet{2020MNRAS.495.3252S} at $z=5$ ({\it left panel}) and $z=4$ ({\it right panel}). Tables for the thick lines bracketing the simulations' functions are available upon request. }
    \label{fig:summary}
\end{figure*}

\section{Summary}
\label{sec:summary}
JWST enables direct comparisons between simulated and observed AGN populations at high redshifts. 
Undoubtedly, the years ahead promise a wealth of additional observations to refine constraints and enhance our ability to improve subgrid models in large-scale cosmological hydrodynamical simulations.
In this paper, we provided a theoretical perspective based on these simulations on the abundance of AGN candidates identified with JWST. 
Our main findings are summarized as follows:

\begin{itemize}
    \item The different subgrid physics employed in cosmological simulations lead to large variations in their BH and AGN populations (Fig.\ref{fig:all_simulations}). 
    Because of their design and calibration, most simulations agree well with high-redshift JWST AGN candidates that lie close to the empirical $M_{\rm BH}-M_{\star}$ scaling relations and fail to produce any BHs with large $M_{\rm BH}/M_{\star}$ ratios in galaxies with $M_{\star}\leqslant 10^{9}\, \rm M_{\odot}$ (the so-called overmassive BHs).
    \item Current uncertainties on BH mass and 
    stellar mass of AGN candidates with potential large $M_{\rm BH}/M_{\star}$ ratios, as well as the possible discrepancies with other independent tracers from the galaxy kinematics (e.g., $\sigma$, $M_{\rm dyn}$), prevent us from drawing strong conclusions about the $M_{\rm BH}-M_{\star}$ diagram at high redshift, a possible relation in that diagram and its evolution with time. 
    \item In the range primarily constrained today with JWST (i.e., $L_{\rm bol}\geqslant 10^{44.5}\, \rm erg/s$), a few simulations produce similar or lower number densities of AGN than the observational constraints (Fig.~\ref{fig:LF_all_simulations}, Fig.~\ref{fig:summary}). As we included all simulated AGN produced in galaxies with $\geqslant 10^{8.5}\, \rm M_{\odot}$ and assumed that none is obscured, our results raise some intriguing questions. 
    When using an empirically-derived X-ray bolometric correction, most of the simulations produce an X-ray (2-10 keV) luminosity function that agrees with observational constraints up to $z\sim5$ coming from X-ray surveys (Fig.~\ref{fig:XrayLF_all_simulations}). The simulations producing the least number of AGN in the luminosity range probed by JWST are also the ones falling short when compared to the bright end of the observed X-ray luminosity functions.
    \item If the JWST observational constraints on the bolometric luminosity function are confirmed, it may be necessary to refine the subgrid physics of certain simulations (e.g., seeding, super-Eddington accretion, SN feedback) to facilitate an overall faster and/or earlier assembly of the BHs.
    \item While the most challenging with JWST, constraining the mass function at $M_{\rm BH}\sim 10^{7}\, \rm M_{\odot}$ or the AGN luminosity function at $L_{\rm bol}\sim 10^{44}\, \rm erg/s$ would yield highly valuable constraints on the subgrid physics of cosmological simulations. These ranges represent points of significant divergence among the simulations, exceeding 1 dex. Such constraints would shed light on the early assembly of massive BHs and, inevitably, the mechanisms responsible for the abundance of AGN.
    
\end{itemize}

For further comparisons between JWST observations and the cosmological simulations, we provide in Fig.~\ref{fig:summary} the summary of the BH mass function and AGN luminosity function produced in simulations at $z=5$ and $z=4$. The thick colored lines bracket the functions and their Poisson errorbars; corresponding Tables are available upon request. Fig.~\ref{fig:summary} also includes new constraints that have been presented in the literature since the submission of this paper \citep{2023arXiv230605448M,2023arXiv230905714G,2023arXiv230801230M,2023arXiv230311946H,2024arXiv240610341A,2024arXiv240919205G,2024ApJ...968...38K}.

\section*{Acknowledgements}
It is my pleasure to thank the PIs and other colleagues who have shared the data from their cosmological simulations with the community and myself.
I warmly thank Marta Volonteri for her helpful comments on this manuscript and inspiring discussions, as always. 
I also thank Ektoras Pouliasis for fruitful discussions regarding observational constraints on the X-ray luminosity function and for providing additional data. I acknowledge funding from the Swiss SNSF Starting Grant (218032).

\section*{Data Availability}
Illustris and TNG data are available at: https://www.illustris-project.org, https://www.tng-project.org. EAGLE data can be accessed at icc.dur.ac.uk/Eagle, while SIMBA data is available at simba.roe.ac.uk.
Data from BlueTides and Astrid can be found at: https://bluetides-portal.psc.edu/simulation/1/, https://astrid-portal.psc.edu/simulation/1/.
Horizon-AGN is not public (but the data can be shared upon request); some catalogs are available at: https://www.horizon-simulation.org/data.html. 
Catalogs including galaxy and BH properties can be shared upon request for some simulations.

\bibliographystyle{mnras}
\bibliography{biblio} 
\label{lastpage}
\end{document}